\documentclass{rsproca_trinh}
\usepackage{microtype}

\usepackage{amsmath,amssymb,amsfonts,upref,endnotes}
\usepackage{amsthm}

\usepackage{pdflscape}
\usepackage{afterpage}
\usepackage{rotating}
\usepackage{fancyhdr}
\fancypagestyle{lscape}{%
\fancyhf{} 
\fancyfoot[LE]{}
\fancyfoot[LO] {}
 
}

\usepackage[sort, numbers]{natbib}

\newtheorem{result}{\bf Result}[section]

\pdfmapfile{+rsfso.map}
\DeclareSymbolFont{rsfso}{U}{rsfso}{m}{n}
\DeclareSymbolFontAlphabet{\mathscr}{rsfso}

\newcommand*{\ep}{\epsilon}

\renewcommand*{\i}{\mathrm{i}}
\newcommand*{\im}{\mathrm{i}}
\newcommand*{\e}{\mathrm{e}}
\newcommand*{\Oh}{\mathcal{O}}
\newcommand*{\bq}{\bar{q}}

\newcommand*{\bt}{\bar{\theta}}

\renewcommand*{\H}{\mathscr{H}}
\newcommand*{\bH}{\hat{\H}}
\renewcommand*{\Re}{\operatorname{Re}}
\renewcommand*{\Im}{\operatorname{Im}}
\newcommand*{\Arg}{\operatorname{Arg}}
\renewcommand*{\j}{\mathrm{j}}

\newcommand*{\Eb}{\mathcal{E}_\mathrm{bern}}

\newcommand*{\ws}{\phi_\text{start}}
\newcommand*{\Ate}{\text{A}}
\newcommand*{\Bte}{\text{B}}
\newcommand*{\Cte}{\text{C}}

\newcommand*{\Iend}{I_\mathrm{endpt}}
\newcommand*{\Iexp}{I_\mathrm{exp}}
\newcommand*{\qexp}{q_\mathrm{exp}}

\newcommand*{\de}{\operatorname{d\!}{}} 
\newcommand{\dd}[2]{\frac{\de#1}{\de#2}}

\usepackage{mathabx}
\usepackage{esint} 
\def\Xint#1{\mathchoice
   {\XXint\displaystyle\textstyle{#1}}%
   {\XXint\textstyle\scriptstyle{#1}}%
   {\XXint\scriptstyle\scriptscriptstyle{#1}}%
   {\XXint\scriptscriptstyle\scriptscriptstyle{#1}}%
   \!\int}
\def\XXint#1#2#3{{\setbox0=\hbox{$#1{#2#3}{\int}$}
     \vcenter{\hbox{$#2#3$}}\kern-.5\wd0}}

\def\YYint#1#2#3{{\setbox0=\hbox{$#1{#2#3}{\int}$}
     \vcenter{\hbox{\scalebox{1}[-1]{$#2#3$}}}\kern-.5\wd0}}

\def\dashint{\Xint-}
\def\cint{\Xint\righttoleftarrow}

\begin{document}

\title{A topological study of gravity waves generated by moving bodies using the method of steepest descents}

\author{
Philippe H. Trinh}

\address{Oxford Centre for Industrial and Applied Mathematics, Mathematical Institute, University of Oxford, \\Oxford OX2 6GG, UK}

\subject{fluid mechanics, wave motion, applied mathematics}


\corres{P.H. Trinh\\
\email{trinh@maths.ox.ac.uk}}

\begin{abstract}
The standard analytical approach for studying gravity free-surface waves generated by a moving body often relies upon a linearization of the physical geometry, where the body is considered asymptotically small in one or several of its dimensions. In this paper, a methodology that avoids any such geometrical simplification is presented for the case of flows at low speeds. The approach is made possible through a reduction of the water-wave equations to a complex-valued integral equation that can be studied using the method of steepest descents. The main result is a theory that establishes a correspondence between a given physical flow geometry, with the topology of the Riemann surface formed by the steepest descent paths. Then, when a geometrical feature of the body is modified, a corresponding change to the Riemann surface is observed, and the resultant effects to the water waves can be derived. This visual procedure is demonstrated for the case of two-dimensional free-surface flow past a surface-piercing ship and over an angled step in a channel.
\end{abstract}



\begin{fmtext}
\section{Introduction}

Let us consider the problem of determining the surface gravity waves generated by a body moving in a two-dimensional potential fluid. At the free surface, $y = \eta(x)$, Bernoulli's equation requires that 
\begin{equation} \label{berntest}
  \frac{1}{2} q^2 + g\eta = \text{const.},
\end{equation}
{where $q$ is the fluid speed and $g$ is the gravitational parameter. The nonlinearity of \eqref{berntest} forms the primary difficulty of analysis; in order to make any sort of progress, the equation must usually be linearized. As explained by Tuck~\cite{tuck_1964}, this linearization will typically involve making one of two possible assumptions.}
\end{fmtext}


\maketitle

In the first, the relevant flow quantities are expressed as a series expansion in powers of a geometric parameter such as $\delta = \text{body size/depth of submergence}$. For instance, one might assume that at leading order, the object is asymptotically thin or streamline in one or several of its dimensions. As particular examples of note, a ship can be approximated by a single point source (as studied by Lord Kelvin~\cite{thomson_1906}) or a series of point sources along its centre plane (as studied by Michell~\cite{mitchell_1898}). Alternatively, but also consistent with the limit $\delta \to 0$, is the assumption that the body is submerged at a large depth compared to its own dimensions. Then, for such an approach, the leading-order solution describes flow past the body in an infinite fluid without a free surface. 

However, approximation schemes dependent on taking the limit of some geometric parameter can be difficult to justify formally as boundary conditions are applied at approximate locations rather than exact locations~\cite{tuck_1964}. Thus for instance, flow past a circular cylinder can be approximated by flow past a dipole, but a finite-order approximation (in powers of the cylinder radius) can never yield a closed streamline around the body. Even beyond the inability to satisfy the exact body conditions, another limitation of such approaches is that they say very little about the situation of flow around nonlinear or bluff bodies; this is a crucial limitation for our purposes.

The second linearization of \eqref{berntest} is more suitable for developing a theory of thick or bluff bodies. In this approach, the relevant small parameter is typically related to the Froude number, written as $\ep = \textrm{inertial/gravitational forces}$. As $\ep \to 0$, the body is assumed to move at low speeds, and at leading-order, the free surface is entirely flat. The great advantage of this approach is that there is no geometrical restriction placed on the body; thus the associated boundary conditions are satisfied exactly to all orders. The challenge, however, is that the limit $\ep \to 0$ is singular. The free-surface waves are exponentially small and their amplitudes scale as $\e^{-\text{const.}/\ep}$. Thus they are said to be \emph{beyond-all-orders} of any regular expansion in powers of $\ep$. The inability to detect the surface waves using standardized methods in perturbation theory is known as Ogilvie's \emph{Low-Froude paradox}~\cite{,tulin_2005_reminiscences_and}, based on Francis Ogilvie's seminal observations in 1968~\cite{ogilvie_1968_wave_resistance:}.

In the fifty years that followed Ogilvie's original work, the low-Froude paradox was eventually resolved through the combined efforts of many researchers. We mention, for example, works by Dagan \& Tulin~\cite{dagan_1972_two-dimensional_free-surface}, Vanden-Breock \& Tuck~\cite{vanden-broeck_1977_computation_of}, and Ogilvie \& Chen~\cite{ogilvie_1982_water_waves}, but a more comprehensive review can be found in \cite{trinh_tulinmodel_paper,tulin_2005_reminiscences_and}. Chapman \& Vanden-Broeck~\cite{chapman_2006_exponential_asymptotics} were the first to leverage modern techniques of exponential asymptotics to the study of the low-Froude flows---in prior work, without such techniques, the derived surface quantities were often asymptotically inconsistent. The difficulty, though, in applying such techniques is their relative complexity; these methods requires notions of estimating asymptotic divergence, optimal truncation, inner-and-outer boundary layer matching, and the Stokes Phenomenon. (see \emph{e.g.} works by Boyd~\cite{boyd_1999_the_devils}, Costin~\cite{costin_book}, Olde Daalhuis \emph{et al.}~\cite{olde-daalhuis_1995_stokes_phenomenon}, Berry and Howls~\cite{berry_1991_asymptotics_superasymptotics}).

We ask: \emph{is there a more intuitive approach to the study of such low-Froude flows?}

In this paper, we shall demonstrate that, for the particular case of gravity-driven flow, there exists a powerful and visual methodology for the study of wave-body interactions. This approach is made possible through a reduction of the water wave equations to a single first-order complex-valued differential equation that can be integrated. In fact, there is a rich history that underlies the development of the methodology, as introduced in the companion paper~\cite{trinh_tulinmodel_paper}, and originating from the works of Tulin~\cite{tulin_1982_an_exact} and Tuck~\cite{tuck_1991_waveless_solutions}. Before embarking on the present paper, we highly recommend a reading of the historical significance of the problem, as presented in \cite{trinh_tulinmodel_paper}. We first provide a brief review of the main results of the companion paper. 

\section{The reduced water-wave equations} \label{sec:mathintro}

\begin{figure}[p]\centering
\includegraphics[scale=1]{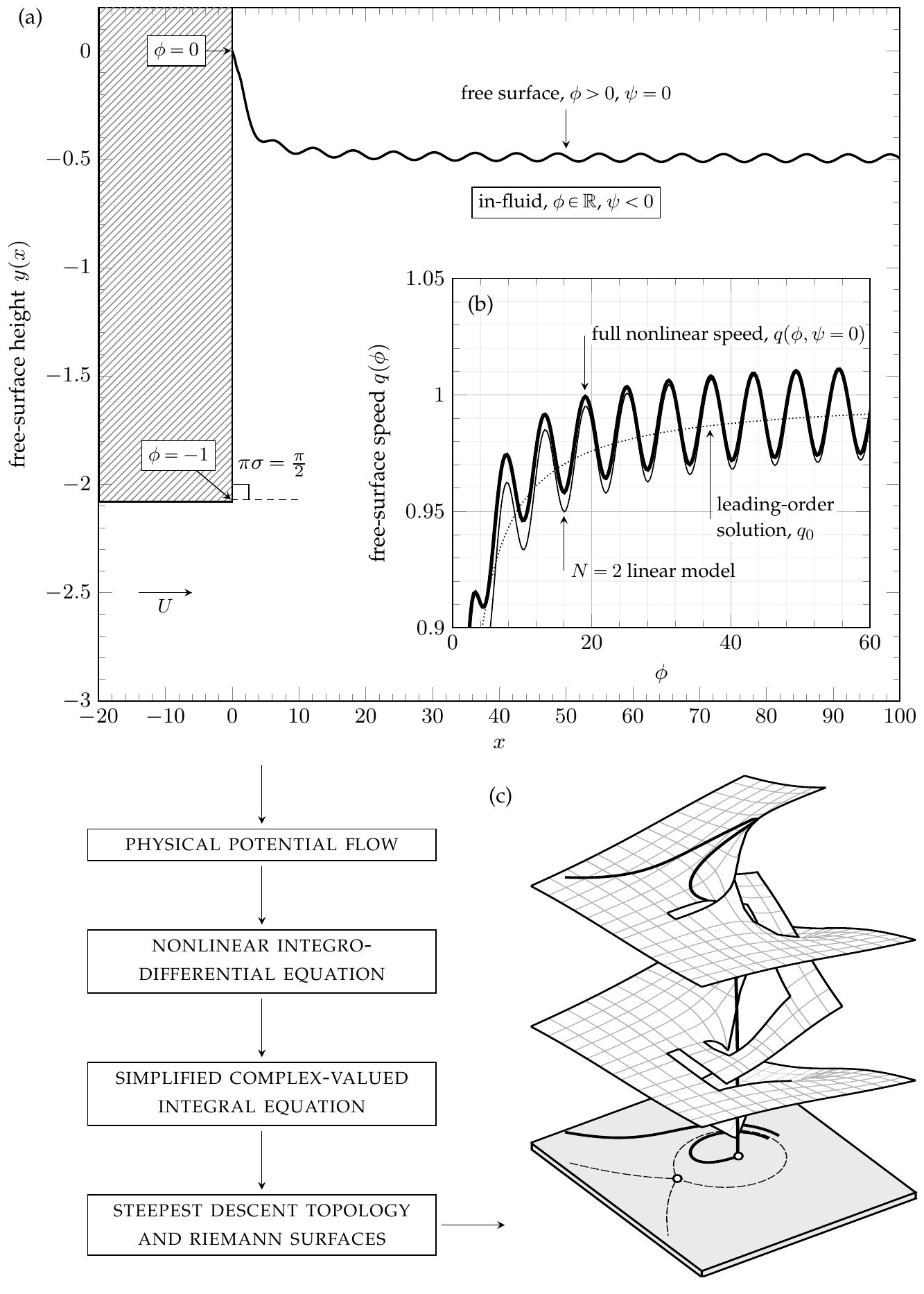}
\caption{The main goal of this paper is to establish a correspondence between the physical problem of flow past a moving body (a) with the topology of a Riemann surface (c). These illustrations correspond to the rectangular ship \eqref{theta_ship} with $\sigma = 1/2$. The surface profile, $y(x)$, in (a) corresponds to the numerical solution of the full nonlinear equations \eqref{govship} at $\ep = 1.0$. The profiles in (b) correspond to the surface speed $q(\phi)$, as a function of the potential $\phi$; as seen, there is close agreement with the simplified models. The Riemann surface (c) is explained in Fig.~\ref{fig:surfship}. \label{fig:outline}}
\end{figure}

For the moment, let us consider potential free-surface flow past the stern of a surface-piercing ship, modeled as the two-dimensional semi-infinite geometry shown in Fig.~\ref{fig:outline}(a). The physical coordinates are denoted by $z = x + \im y$ and the complex potential by $w = \phi + \im \psi$, for velocity potential $\phi$ and streamfunction $\psi$. Then, our task is to solve for the speed, $q(\phi, \psi)$, and streamline angle, $\theta(\phi, \psi)$, on the free surface, where $\phi > 0$ and $\psi = 0$. The governing water-wave equations are reviewed in Appendix~\ref{sec:gov}, and are given in non-dimensional form by
\begin{equation} \label{govship}
\ep q^2 \dd{q}{\phi} + \sin \theta = 0 \quad \text{and} \quad
\log q = \frac{1}{\pi} \int_{-\infty}^0 \frac{\theta(\varphi)}{\varphi - \phi} \, \de{\varphi} + \H\theta(\phi) \quad \text{on $\psi = 0$}.
\end{equation}
Here, we have introduced an operator for a principal-value integral known as the Hilbert transform. It is defined by 
\begin{equation}
  \H\theta(\phi) = \H[\theta] = \frac{1}{\pi} \dashint_{0}^\infty \frac{\theta(\varphi)}{\varphi - \phi} \, \de{\varphi}.
\end{equation}

The first equation in \eqref{govship} is the differentiated form of Bernoulli's condition [\emph{c.f.} \eqref{berntest}], with $\ep = \textrm{Fr}^2 = U^2/(gL)$ being the square of the Froude number for upstream speed $U$, gravity $g$, and geometrical length scale $L$. The second is a boundary-integral equation, written so as to split the integral on the solid body $(\phi < 0)$ where $\theta$ is assumed to be known, with the Hilbert transform of $\theta$ over the unknown free surface $(\phi > 0)$. For the surface-piercing configuration, we also specify that the free surface attaches to the body at a stagnation point, $q(0) = 0$. 

For a semi-infinite ship with the single stern-face of angle $\pi \sigma$ shown in Fig.~\ref{fig:outline}, the only length scale is the distance from the stagnation point to the corner. Thus, we may non-dimensionalize the problem so that the corner lies at $\phi = -1$, and the body can be specified as
\begin{equation} \label{theta_ship}
  \theta_\textrm{ship} = \begin{cases}
  0 & \text{for $\phi \in (-\infty, -1)$}, \\ 
  \pi\sigma & \text{for $\phi \in (-1, 0)$}.
  \end{cases}
\end{equation}
The above body shape is then used in \eqref{govship}, and the two equations can be numerically solved using the finite-difference schemes discussed in \emph{e.g.} \cite{trinh_2011_do_waveless,vb_book}. In Fig.~\ref{fig:outline}(a), we have plotted the physical free surface height, $y(x)$, obtained by integrating $\dd{z}{w} = \frac{\e^{\i\theta}}{q}$ once $q$ and $\theta$ are known.

A typical numerical solution for the case of a rectangular stern, $\sigma = 1/2$, is shown in Fig.~\ref{fig:outline}(a) at $\ep = 1$. Although the two-dimensional semi-infinite geometry of \eqref{theta_ship} may seem rather idealized, it presents a canonical model of wave-structure interaction with a variety of numerical and analytical challenges. These two-dimensional hull shapes have been considered in \emph{e.g.} the works of \cite{vanden-broeck_1978_divergent_low-froude-number,trinh_2011_do_waveless}, but similar considerations can be extended to ships with multiple corners \cite{trinh_2014_the_wake}, or ships with smooth or bulbous profiles \cite{farrow_1995_further_studies}.

Now the major difficulty in solving \eqref{govship} is that the Hilbert transform is a global operator, and thus it requires the very values of $\theta$ that we seek. Earlier, Tuck~\cite{tuck_1991_waveless_solutions} had proposed the idea that toy models of certain wave-structure problems could be derived by ignoring or replacing the Hilbert transform, and changing \eqref{govship} to a local problem of initial-value type. In the companion paper \cite{trinh_tulinmodel_paper}, such a reduction was shown to be formally possible in the limit $\ep \to 0$.

The argument is as follows. Let us expand the solution of \eqref{govship} as a truncated regular series expansion and an error term,
\begin{equation} \label{qtseries}
  q = \Bigl[q_0 + \ep q_1 + \ldots + \ep^{N-1} q_{N-1} \Bigr] + \bq \quad \text{and} \quad
  \theta = \Bigl[\theta_0 + \ep q_1 + \ldots + \ep^{N-1} \theta_{N-1} \Bigr] + \bt.
\end{equation}
It follows from setting $\ep = 0$ in \eqref{govship} that at leading order, $\theta_0 = 0$, and the free-surface is entirely flat. Effectively, the surface has been replaced by a rigid plane, and the solution is known as the \emph{double-body flow}. In this case, the leading-order speed is given by 
\begin{equation} \label{eq:q0ship}
  q_0 = \left(\frac{\phi}{\phi + 1}\right)^\sigma.
\end{equation}
As seen in Fig.~\ref{fig:outline}, the leading-order solution fails to capture the wave phenomena. In fact, the situation is much worse, and it can be argued through the study of higher-order terms in \eqref{qtseries}, that none of the terms, $q_n$ and $\theta_n$, will contain an oscillatory component at any order. Indeed, this is Ogilvie's low-speed paradox~\cite{ogilvie_1968_wave_resistance:}. 

Let us take $q$ and $\theta$, defined on the free surface, $\phi > 0$, and analytically continue the solutions to complex values of $\phi$. Relabeling $\phi \mapsto w\in\mathbb{C}$, the system \eqref{govship} can be reduced to the following complex-valued differential equation:
\begin{equation} \label{qbeqnship}
\ep \dd{\bq}{w} + \biggl[ \dd{\chi}{w} + \ep \dd{P_1}{w}\biggr]\bq 
\sim R(w; \bH[\bt]) \quad \text{where} \quad \bH\bt(w) = \bH[\bt] = \int_0^\infty \frac{\bt(\varphi)}{\varphi - w} \, \de{\varphi}.
\end{equation}
The derivation of \eqref{qbeqnship} is reviewed in Appendix~\ref{sec:reduce}. The idea of extending the governing system \eqref{govship} to complex values may seem odd, but the rationale can be seen \emph{a posteriori}.

Above, we have assumed that $w$ lies in the upper half-plane (an analogous derivation can be done for the lower half-plane). We have also introduced $\bH$ for the complexified Hilbert transform, which is non-singular for $w$ off the real axis. The two functions, $\chi$ and $P_1$, \eqref{qbeqnship} are known, typically in closed form, while $R$ may be approximated; we will specify these quantities later. Integrating the equation, and taking the limit whereby we return to the physical free surface, $w \to \phi > 0$, we obtain the result of 
\begin{equation} \label{myint}
\bq(\phi) \sim \left[\frac{\e^{-P_1(\phi)}}{\ep}\right] \left[ \int_{\phi_\text{start}}^\phi R(\varphi; \bH[\bt]) \e^{P_1(\varphi)} \, \e^{\chi(\varphi)/\ep}\, \de{\varphi}  \right] \e^{-\chi(\phi)/\ep}.
\end{equation}
In the example of flow past a surface-piercing ship, we take ${\phi_\text{start}} \to 0$ in order to account for condition of a stagnation point, $\bq(0) = 0$, at the edge of the hull.

There are many subtleties to the reduction process that takes the full system \eqref{govship} to the first-order problem \eqref{qbeqnship}, and then again to the integral \eqref{myint}. Indeed, a rigorous explanation of why this reduction is valid forms the basis of the companion work~\cite{trinh_tulinmodel_paper}. For now, let us take \eqref{myint} for granted, and seek an approximation of the integral. 

In the limit $\ep \to 0$, the initial contour of $\varphi \in [0, \phi]$ can be defomred along paths of steepest descent. Once done, an asymptotic expansion of the integral is derived by consideration of dominant contributions from the endpoints, as well as any relevant saddle or critical points of the integrand. In \cite{trinh_tulinmodel_paper}, it was shown that the endpoint contributions would lead to a re-expansion of higher-order algebraic terms from \eqref{qtseries}, while the water waves were caused due to contributions \emph{away} from the positive real axis. Therefore $\bH[\bt]$, which only depends on the free surface values of $\bt$, can be ignored at leading order. This subtle reduction process explains why reduced wave models (such as the ones reviewed in \cite{trinh_tulinmodel_paper}) remove or substitute alternatives for the Hilbert transform. 

Let $\qexp$ be the exponentially waves that arise from the integral \eqref{myint} according to this process. The free surface will then be approximated by 
\begin{equation} \label{q0plus}
  q \sim \Bigl[q_0 + \ep q_1 + \Oh(\ep^2)\Bigr] + 2 \Re(\qexp),
\end{equation}
which includes both the wave-free flow (the square-bracketed terms) and the surface waves. Note that we are required to add both $\qexp$ from \eqref{myint} and also its complex conjugate, due to an analogous argument for the analytic continuation of \eqref{qbeqnship} into the lower half-plane, and subsequent application of the Schwartz reflection principle. Combining these two contributions then leads to $2 \Re(\qexp)$. 

Let us be more concrete and provide a numerical example. Suppose we return to the differential equation \eqref{qbeqnship} and, having been sufficiently convinced by the above argument, we ignore the $\bH[\bt]$ contributions to $R$. Using the initial condition $\bq(0) = 0$, the values \eqref{chi} and \eqref{P1} for $\chi$ and $P_1$, and the approximation \eqref{Rtest} for $R$, the differential equation is solved as an intial-value problem. The result, \eqref{q0plus}, is shown in Fig.~\ref{fig:outline}(b). The agreement with the full nonlinear solution is superb, even at the moderate value of $\ep = 1$. In \cite{trinh_tulinmodel_paper}, a careful check of numerical vs. asymptotic results as $\ep \to 0$ verifies these claims properly. 

Now we arrive at the current aims of this paper. Previously in \cite{trinh_tulinmodel_paper}, the asymptotic approximations of the integral \eqref{myint} were made based on the general assumption that the initial contour of integration $\varphi \in [\phi_\text{start}, \phi]$, could be deformed for any given body geometry, and that such a deformation would include the key critical points generating the water waves. In actuality, the topology of the steepest descent paths may be rather complicated. Different body geometries yield different contributing singularities; or perhaps for some bodies, there may be branch cuts that complicate the deformation procedure.

Thus, our main goal is to demonstrate, through the use of several representative examples, the correspondence between the physical flow and the topology of steepest descent paths. The power of this approach is two-fold. First, it provides an analytical methodology for determining a wide range of wave-structure interaction problems. Second, it provides an intuitive approach to visualizing the effects of changing particular geometrical properties. We will see, for example, how the complex-valued topology of a surface-piercing ship differs from that of channel flow, or how flow past a shallow angled step differs from flow past a rectangular step. This correspondence process is illustrated in Fig.~\ref{fig:outline}(c). 

\section{Computation of the Riemann surfaces} \label{sec:riemann}

Let us now explain how the Riemann surfaces connected to the steepest descent paths are computed. A general reference for the method of steepest descents can be found in \emph{e.g.} \cite[Chap. 7]{bleistein_1975_asymptotic_expansions}. 

We first re-write the integral \eqref{myint} as 
\begin{equation} \label{Iint}
  I(\phi) = \int_{\ws}^{\phi} f(\varphi) \e^{\chi(\varphi)/\ep} \, \de{\varphi} \quad \text{where} \quad
  f(\varphi) = R(\varphi; \bH[\bt]) \e^{P_1(\varphi)},
\end{equation}
and $\chi$, a quantity known as the \emph{singulant}, is given by \eqref{chi}, or repeated here as
\begin{equation} \label{eq:chimain}
\chi(w) = \int_{w_0}^w \frac{\im \j}{q_0^3(\varphi)} \, \de{\varphi}.
\end{equation}
In the definition of $\chi$, we have introduced the sign constant, $\j$. The constant takes the value $\j = 1$ for surface-piercing flows, and $\j = -1$ for channel flows. This distinction is necessary due to the difference in surface flow direction relative to the solid body, as explained in \eqref{jsign}. The function $q_0$ is the double-body solution discussed in \S\ref{sec:mathintro}, and encodes the information for the geometry of the moving body. Any choice for the initial point of integration, $w_0$, is valid so long as the integral is defined; a selection will be compensated by the numerical pre-factors associated with $f(\varphi)$. We will discuss specifics of the $\chi$ function later. 

Returning to \eqref{Iint}, the contour is initially along the real axis, $\varphi \in [\ws, \phi]$, but once it is deformed along the paths of steepest descent, this would yield the general decomposition of 
\begin{equation}
  I(\phi) \sim \Iend(\phi) + \Iexp.
\end{equation}
Thus the integral is approximated by the endpoint contributions, $\Iend$, from $\varphi = \{\ws, \phi\}$, and further contributions, $\Iexp$, from critical points (saddle points, poles, or branch points), presumed to correspond to the surface waves. The paths of steepest descent are given by level sets of $\Im(\chi)$, and thus the criterion for determining whether $w$ lies on the steepest descent path from $w_0$ is
\begin{equation} \label{eq:steepcrit}
  \Im[\chi(w)] = \Im[ \chi(w_0)] \quad \text{and} \quad \Re[\chi(w)] \leq \Re[ \chi(w_0)].
\end{equation}
The first condition ensures that $w$ is on a path of steepest descent or ascent; the second ensures that it is, in fact, a path of steepest descent. 

In practice, however, calculating $\chi(w)$ for different values of $w$ is not necessarily straightforward, as depending on the body function, $q_0$, $\chi$ may possess a complicated branch structure. A similar issue was encountered in the development of asymptotic solutions to the problem of thin-film rupture~\cite{chapman_2013_exponential_asymptotics}. In order to explain the terminology for the branch structure, let us return to the example ship flow of \S\ref{sec:mathintro}. Substituting $q_0$ from \eqref{eq:q0ship} into \eqref{eq:chimain}, and choosing the initial point of integration, $w_0 = -1$, to correspond to the corner of the ship, we have
\begin{equation} \label{eq:chi_ship}
 \chi(w)  = \chi_{(k_1, k_2)} = \i \int_{-1}^w \left(\frac{\varphi + 1}{\varphi}\right)^{3\sigma} \, \de{\varphi}.
\end{equation}
To define a single-valued function for the above integral, we must specify the branches and branch cuts associated with the two critical points at $\varphi = W_i \in \{-1, 0\}$. Each single-valued specification of $\chi$ is called a \emph{branch} (the output or range), and each branch is defined on a \emph{Riemann sheet}, $S \subset \mathbb{C}\setminus\{-1, 0\}$ of the complex plane (the input or domain). 

However, as $w$ varies in the complex plane, the contour of integration may cross a branch cut; here, the single-valued specification of $\chi$ encounters a discontinuity. For our purposes, we wish to continue the trajectory of the steepest descent contours of \eqref{Iint} beyond such discontinuities. In other words, when a branch cut is traversed, we will move onto the adjacent Riemann sheet, and take the values of $\chi$ from the next branch. The collection of Riemann sheets is known as the \emph{Riemann surface} for $\chi$. It is this notion of a Riemann surface, represented by copies of the complex plane attached to one-another, that allows us to unambiguously define a multi-valued function for $\chi$. We refer readers to \cite[Chap.~2]{alekseev_book} or \cite[Chap.~7]{wegert_book} for more background on such complex-variable terminology.


Continuing with our ship example in \eqref{eq:chi_ship}, let us choose the branch cuts so as to lie along the rays of constant angle from each branch point, $w = W_i$. Typically, we will choose them so as to lie parallel to one of the coordinate axes. Let $k_i\in \mathbb{Z}$ be the winding number around the point $W_i$. That is, $k_i$ marks the number of rotations associated with a given analytic continuation around the point $W_i$, with positive and negative integers for counterclockwise and clockwise rotations, respectively. 

Notationally, we will denote each distinct branch of $\chi$ by the pair $(k_1, k_2)$, indicating the number of revolutions required in order to reach the current branch. For example, consider the surface-piercing ship with branch points at $W_1 = -1$ (the corner) and $W_2 = 0$ (the stagnation point). Three illustrative branches are labeled as
\begin{equation} \label{chiexample}
 \chi_{(0,0)}, \quad  
 \chi_{(0,1)}, \quad \text{and} \quad
 \chi_{(0,1) \to (1,1)}.
\end{equation} 
The first, $\chi_{(0,0)}$, is the principal branch obtained by integrating \eqref{eq:chi_ship} from $\varphi = -1$ to $\varphi = w$ without crossing a branch cut. The second, $\chi_{(0,1)}$ is found by integrating from $\varphi = -1$, then circling the stagnation point, $W_2$, in the positive sense, and arriving at $\varphi = w$. The third, $\chi_{(0,1) \to (1,1)}$, begins from the previous branch, and then performs an additional rotation about the corner, $W_1$. When it may be ambiguous, we use the arrow notation to clarify the order in which branch cuts are crossed.

A three-dimensional cross-section of the Riemann surface can be visualized by plotting $(\Re w, \Im w, \Re \chi)$ using $\chi_{(0,0)}$ and its adjacent sheets. In general, these surfaces will self-intersect within the graphical representation but they are otherwise free of discontinuities. For similar visualizations of Riemann surfaces, see \cite[p.~209]{alekseev_book} or \cite[Fig.~7.2]{wegert_book}. 

The Riemann surfaces that appear in this paper, such as the one in Fig.~\ref{fig:outline}, are graphed from the numerical integration of $\chi$ using the trapezoid rule. In such computations, we have carefully defined the locations of the branch cuts, and implemented a numerical scheme allowing the integration path to smoothly switch between the individual branch sheets. Both interactive plots of the Riemann surfaces, and also the data for the contour plots appearing in this paper can be found within the electronic supplementary material.

\section{The paths of steepest descent} \label{sec:steep}

We are now ready to study the steepest descent paths associated with the integral \eqref{eq:chimain}. We have chosen to focus on three representative geometries: the previous rectangular semi-infinite ship in infinite depth [in \S\ref{sec:steep}\ref{sec:ship}], a rectangular step in a channel [in \S\ref{sec:steep}\ref{sec:step}], and an angled step in a channel [in \S\ref{sec:steep}\ref{sec:stepangle}]. The methods we present can be extended to cover many other classes of moving bodies, notably those for which a boundary integral formulation exists. This includes, for instance, models of an air jet stream blowing on stationary water~\cite{grundy_1987_waves_on}, flow over a submerged source or sink~\cite{lustri_2013_exponential_asymptotics}, and flows past sluice gates or surf boards~\cite{binder_2005_free_surface, binder_2007_the_effect}. 



\subsection{The rectangular ship in infinite depth} \label{sec:ship}

Let us return to the case of flow past a semi-infinite rectangular stern. The geometry is specified by $\theta_\text{ship}$ in \eqref{theta_ship} with $\sigma = 1/2$, and the leading-order speed is given by \eqref{eq:q0ship}. Our task is to approximate the integral \eqref{Iint}, which involves the singulant function \eqref{eq:chi_ship}. The critical points in the $w$ or $\varphi$-plane include the corner $w = -1$ and the stagnation point $w = 0$, and the branch cuts from both are taken along the positive real axis.

In the integral \eqref{Iint}, we would like to take the initial point of integration to be the stagnation point, $\ws = 0$, so as to make use of the boundary condition $\bq(0) = 0$. However, because the origin is a degenerate point, it is easier to first consider the descents paths for the integral
\begin{equation} \label{eq:shipint_again}
  I(\phi) = \int_\Ate^\phi f(\varphi) \e^{\chi(\varphi)/\ep} \, \de{\varphi},
\end{equation}
where $\varphi = \Ate$ is a small and positive number. Also, it can be verified, through examination of the pre-factor $f(\varphi)$, that the steepest descent contours will only depend on $\chi$. Below, we will use $w$ interchangeably with $\varphi$ (the dummy variable). 

\begin{figure}[htb] \centering
\includegraphics[width=1.0\textwidth]{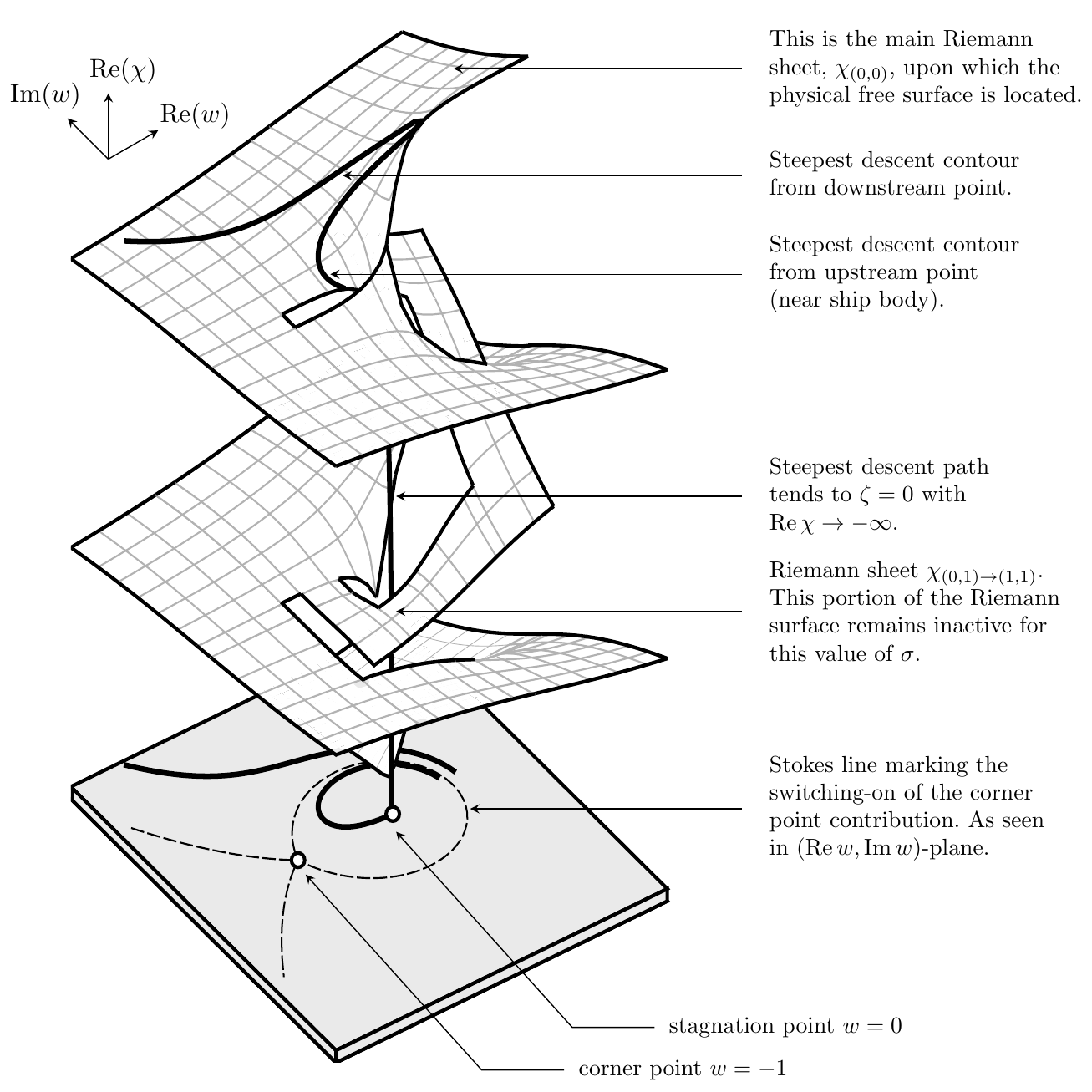}
\caption{
Three-dimensional representation of the Riemann surface of $\chi$ for the rectangular ship geometry with $\sigma = 1/2$, shown in the $(\Re w, \Im w, \Re \chi)$ plane. The two thick contours correspond to steepest descent trajectories for an upstream (near the stagnation point) and downstream point on the free surface. The lower plane shows the projection of curves onto the $(\Re w, \Im w)$-plane. The surface is generated from numerical integration of $\chi$. Note that $\varphi$ is the dummy variable for integration in the $w$-plane.\label{fig:surfship}}
\end{figure}

Using the methodology outlined in \S\ref{sec:riemann}, we generate the Riemann surface associated with $\chi$, and the result is shown in Fig.~\ref{fig:surfship}. The visualized surface is composed of the three listed branches in \eqref{chiexample}. However, as it turns out, in the rectangular geometry, the adjacent Riemann sheets due to the corner point remain inactive, and are untouched by the relevant steepest descent paths. This will not always be the case for all geometries and through all values of the angle $\pi\sigma$. In both the examples of a rectangular and angled step in \S\ref{sec:steep}\ref{sec:step} and \S\ref{sec:steep}\ref{sec:stepangle}, the steepest descent contours will indeed traverse onto adjacent Riemann sheets. 

We continue to examine Fig.~\ref{fig:surfship}, but now turn our attention to the planar projection, shown in grey, where several constant phase curves, $\Im(\chi) = \text{constant}$, are projected down from the Riemann surface. The dashed lines are constant phase lines emerging from the corner point, $w = -1$, with one set of lines, on the right, forming a homoclinic, and the other, on the left, forming an open triangular region. Setting $w + 1 = r\e^{\i\vartheta}$, we find that the local emergence angles of such paths are
\begin{equation}
  \Arg \chi \sim \frac{5}{2}\vartheta - \pi \Rightarrow
  \begin{cases}
  \Im \chi = 0 & \text{when $\vartheta = \frac{2\pi}{5} n$,} \\
  \Re \chi = 0 & \text{when $\vartheta = \frac{2\pi}{5} \left(n + \frac{1}{2}\right)$,}
  \end{cases}
\end{equation}
where $n \in\mathbb{Z}$. Thus the homoclinic is formed by the two local angles, $\pm 2\pi/5$, while the triangular region is formed by the two local angles, $\pm 4\pi/5$. 

The most important quantity to note is that along the dashed line with $\vartheta = 2\pi/5$, we have $\Re \chi \geq 0$, which indicates that the trajectory along this curve lies uphill of the point $w = -1$ (note that $\chi$ in \eqref{eq:chi_ship} was chosen so that $\chi = 0$ at the corner). We have also plotted two steepest descent trajectories, shown thick. Within the interior of the homoclinic, the steepest descent trajectory forms an arc that tends towards $w = 0$ in a direction tangential to the negative real axis. Exterior of the homoclinic, and to the right of the plane, the trajectory moves to $|w| \to \infty$ along the dashed line with local angle $4\pi/5$. The dashed line, called the \emph{Stokes line}, marks a critical curve across which the steepest descent paths change in dramatic fashion; it is of key importance in what follows.

\subsubsection{Only endpoint contributions} \label{sec:ship_end}

Let us now turn to Fig.~\ref{fig:shipstokes}(a), and consider the integral \eqref{eq:shipint_again} between two points, $\varphi = \Ate$ and $\Bte$, on the upstream side and sufficiently close to the stagnation point at the hull. Consequently, both such points lie within the dashed homoclinic. 

We see that the contour from $\varphi = \Ate$ is deformed in a counterclockwise manner and tends to $\varphi = 0$ where $\Re \chi \to -\infty$. Within this valley, the contour then unwinds clockwise, with $\chi$ increasing until arriving at $\varphi = \Bte$. Thus, when the integral, $I(\phi)$, is evaluated at $\phi = \Bte < \Bte^*$, where $\Bte^*$ marks the intersection of the dashed line with the positive real axis, the result of this process yields two dominant endpoint contributions, written loosely as
\begin{equation} \label{Iend}
I(\phi) \sim \Iend,
\end{equation}
the terms of which can be derived by integrating \eqref{eq:shipint_again} by parts. As was shown in \cite{trinh_tulinmodel_paper}, this process will yield the higher-order algebraic corrections in \eqref{qtseries}. 

Now that we have understood the steepest descent paths, we may freely take $\varphi = \text{A} \to 0$, that is, take the initial point of integration in \eqref{eq:shipint_again} to the stagnation point; then, the proper steepest descent deformation for the contour $\varphi \in [0, \Bte]$ involves only the single contour in the upper half-plane connected to $\Bte$. 

\emph{We conclude through this process that when sufficiently close to the body of the ship, there are no free-surface waves in the limit $\ep \to 0$ (to this order of exponential accuracy).}

\subsubsection{Crossing the Stokes line}

\begin{figure}\centering
\includegraphics[scale=1]{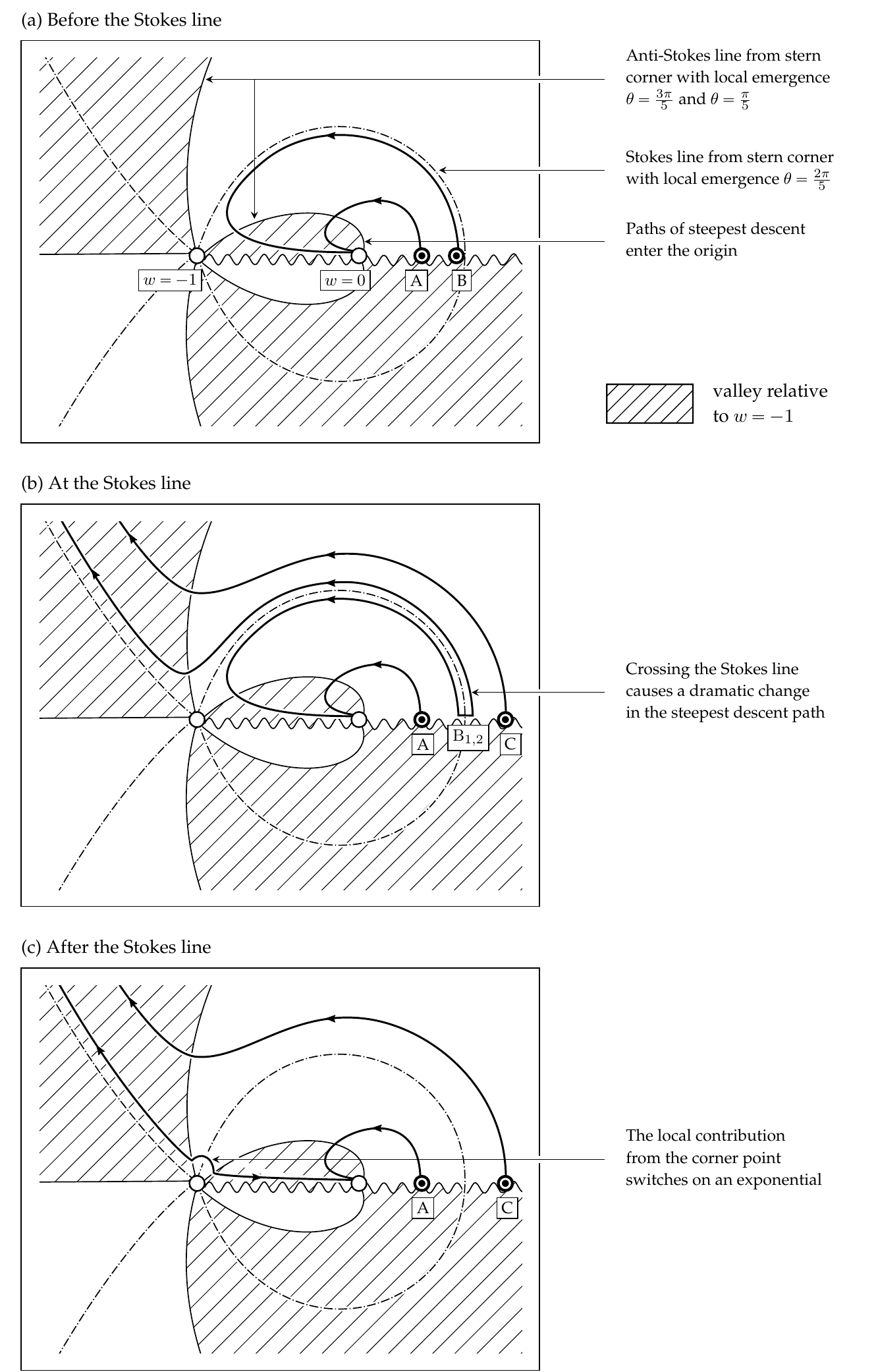}
\caption{Illustration of the steepest descents deformation process for the integral \eqref{eq:shipint_again} and the initial contour from A to B (top) or A to C (middle and bottom). The curves are drawn in the $(\Re w, \Im w$)-plane (for dummy integration variable $\varphi$). Branch cuts are shown as a wavy line. The circles indicate the two points $w = -1$ and $w = 0$. \label{fig:shipstokes}}
\end{figure}

The dashed line in Fig.~\ref{fig:shipstokes}(a) is known as the \emph{Stokes line}, and it marks a critical line across which the analytic continuation of the integral, $I(\phi)$, switches-on a subdominant contribution. The intersection of the Stokes line with the original contour along $\varphi\in[\Ate, \phi]$, is the location across which the deformation process makes a dramatic topological change, and this occurrence is known as the \emph{Stokes Phenomenon}~\cite{trinh_2010_asymptotic_methods_incol}. Let us now consider the method of steepest descents applied to an initial contour in \eqref{eq:shipint_again} beginning at $\varphi = \Ate$ and ending at $\varphi = \phi = \Cte$, which lies outside of the homoclinic region. Thus $\Cte$ has sufficiently far downstream so as to pass the Stokes line. This is shown in Fig.~\ref{fig:shipstokes}(b). Conceptually, it is easier to split the integration range so that
\begin{equation} \label{eq:I_steep_note}
  I(\Cte) = \left(\int_\Ate^{\Bte_1}  + \int_{\Bte_2}^{\Cte}\right) f(\varphi) \e^{\chi(\varphi)/\epsilon} \, \de{\varphi},
\end{equation}
where $\Bte_1$ and $\Bte_2$ are the limiting points just on the interior and exterior of the homoclinic orbit. 

The deformed contour is shown in the figure. The steepest descent contours from $\varphi = \Ate$ and $\Bte_1$ are the same as before, with both curves unraveling within the dashed homoclinic and then connected by the valley at $\varphi = 0$ where $\chi \to -\infty$. However, for the second integral in \eqref{eq:I_steep_note}, the contours from $\varphi = \Bte_2$ and $\Cte$ tend to the valley in the upper left, as $|\varphi| \to \infty$. Since the two integrals that run alongside the top of the homoclinic orbit will sum to zero, this leaves the final deformed contour in Fig.~\ref{fig:shipstokes}(c). Thus, in addition to the endpoint contributions \eqref{Iend}, we have
\begin{equation} \label{IplusIexp}
   I \sim \Iend + \Iexp = \Iend + \cint_{\varphi = -1} f(\varphi) \e^{\chi(\varphi)/\ep} \, \de{\varphi}.
 \end{equation} 
The evaluation of the $\cint$ integral along the steepest descent path depends only on the local properties of the integrand near $\varphi = -1$. This is why we may think of the surface waves as being \emph{generated} by the ship's corner. The exponentially small contribution, corresponding to surface waves, that results from the integral approximation is given in \cite{trinh_tulinmodel_paper}. 

\emph{We conclude that when sufficiently far from the body of the hull, the free-surface waves arise from the switching-on of a saddle-point contribution due to the corner of the ship (the Stokes Phenomenon)}. 

\subsection{The rectangular step in a channel} \label{sec:step}

\begin{figure}[htb] \centering
\includegraphics[width=1.0\textwidth, clip=true, trim=0 4cm 0 0]{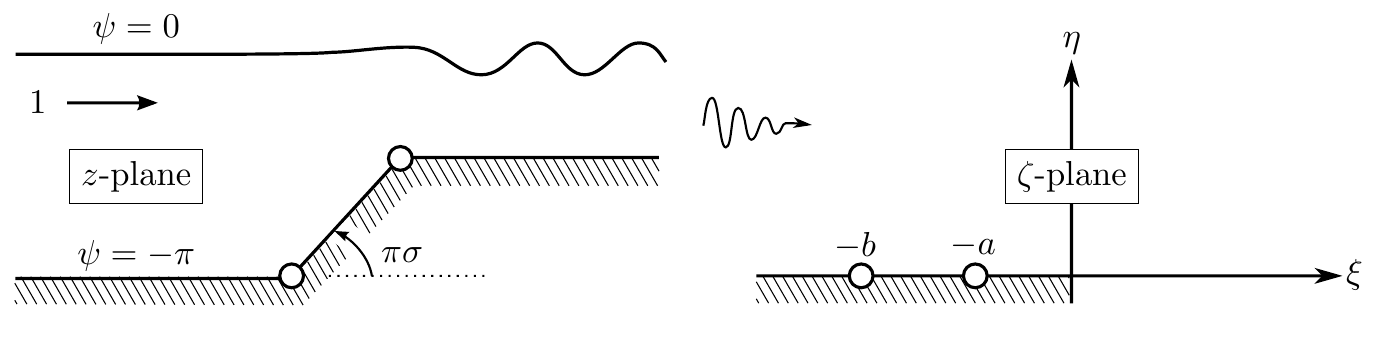}
\caption{Physical (left) and upper-half $\zeta$-plane for the non-dimensional flow over the angled step. The flow is contained within a strap in the complex potential $w$-plane. The map from the strip to the upper-half plane is given by $\zeta = \e^{-w}$. \label{formstep}}
\end{figure}

Consider now the case of an angled step in a channel, as shown in Fig.~\ref{formstep}(a). The problem is non-dimensionalized so that in the potential plane, the free surface is along $\psi = 0$, the channel bottom is along $\psi = -\pi$, and $-\infty < \phi < \infty$. In order to develop the analogous formulae to \eqref{govship}, we map the $w = \phi + \im \psi$ plane to the upper half-plane using $\zeta = \xi + \im\eta$, as shown in Fig.~\ref{formstep}(b) where
\begin{equation} \label{eq:steptrans}
  \zeta = \xi + \im \eta = \e^{-w}.
\end{equation}
Within the $\zeta$-plane, the solid boundary lies along $\xi \leq 0$, the free surface along $\xi \geq 0$, and downstream infinity corresponds to $\zeta \to 0$. 

The development of Bernoulli's equation and the boundary integral equation follow parallel lines, and are given by \eqref{govsys} with $\j = -1$. Having non-dimensionalized with the channel height, we have the freedom to specify the locations of the step's stagnation point, $\zeta = -b$, and corner point, $\zeta = -a$, with $0 < a < b$. The body geometry is described by the angle and speed, 
\begin{equation} \label{qs_step}
  \theta_\textrm{step} = \begin{cases}
  0 & \xi \in (-\infty, -b) \cup (-a, 0)\\ 
  \pi\sigma & \xi \in (-b, -a)
  \end{cases}
\quad \text{and} \quad
  q_s = \left(\frac{\xi + b}{\xi + a}\right)^\sigma = \left(\frac{\e^{-\phi} + b}{\e^{-\phi} + a}\right)^\sigma,
\end{equation}
where the speed follows from \eqref{qsH}. This geometry corresponds to a step-up of angle $\pi\sigma > 0$ to the horizontal, and reversing the flow or allowing a step-down is straightforward. Such step-geometries have been considered by \emph{e.g.} \cite{king_1987_free-surface_flow, chapman_2006_exponential_asymptotics} and, like the semi-infinite ship, they provides a canonical model for free-surface flow in a channel. 

In what follows, we must take care to transform the integration variables from $w$ to $\zeta$. We will abuse notation, and instead of writing $\chi(w) = \chi(-\log\zeta) \equiv X(\zeta)$, we will write $\chi(\zeta)$ for $X(\zeta)$. The analogous integral to \eqref{Iint} and \eqref{eq:chimain}, written now for the $\zeta$-plane is given by
\begin{equation} \label{eq:I_chi_step}
I(\xi) = \int_\Ate^\xi \tilde{f}(t) \e^{\chi(t)/\ep} \, \de{t} \quad \text{with} \quad \chi(\zeta) = -\im \int_{-a}^\zeta \left(\frac{t + a}{t + b}\right)^{3\sigma} \left(-\frac{1}{t}\right) \, \de{t}.
\end{equation}
where $\tilde{f}(t) = -f(\varphi)/\zeta$ using \eqref{Iint} and \eqref{eq:steptrans}. The dummy variable is now $t$, corresponding to integration in the $\zeta$-plane. We have also chosen to centre the singulant so that $\chi(-a) = 0$, and this only changes the numerical pre-factors embedded in $\tilde{f}$. Like the example of the ship, it is convenient to consider the initial point of integration $t = \Ate$ at a finite non-degenerate point along the free-surface, $t \geq 0$, but we will eventually let $t = \Ate \to \infty$, so as to impose the natural upstream boundary condition that requires $\bq \to 0$ as $\zeta = \xi \to \infty$. 

Let us now consider the rectangular step, and set $\sigma = 1/2$. Within integral for the singulant, we choose both branch cuts from $\zeta = -b$ and $\zeta = -a$ to lie in the direction of the positive real axis; due to the combination of powers, this is equivalent to a single branch cut between the two points. There is a further logarithmic branch cut that arises from the pole at $\zeta = 0$, but crossing this cut does not cause any adjustment in the steepest descent paths. 

\begin{figure}\centering
\includegraphics[width=1.0\textwidth]{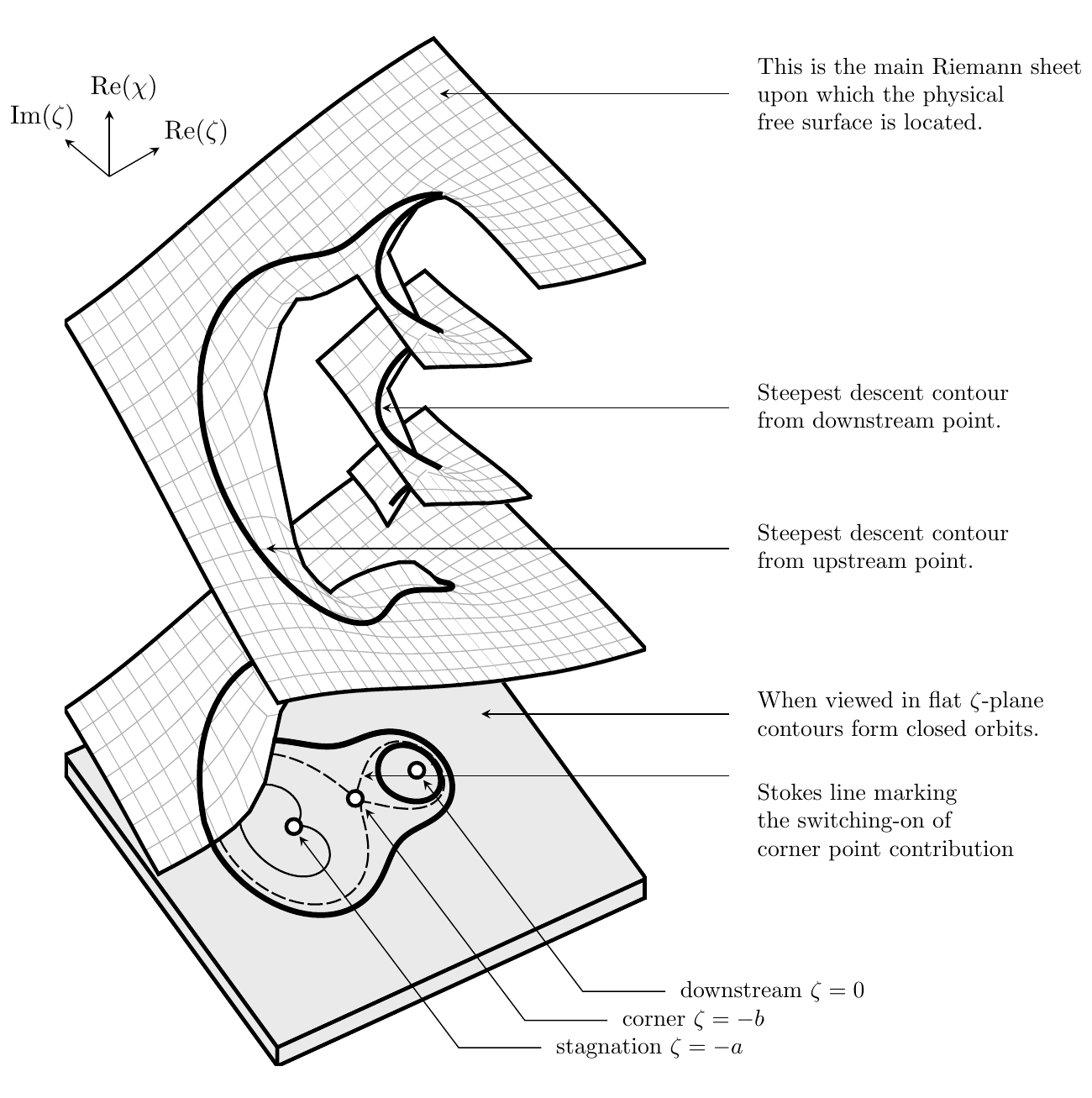}
\caption{A three-dimensional representation of the four-dimensional Riemann surface of $\chi$ for the step geometry with $a = 1$, $b = 2$, and $\sigma = 1/2$, shown in the $(\Re\zeta, \Im\zeta, \Re\chi)$ plane. The two thick contours correspond to steepest descent trajectories for an upstream and downstream point on the free surface. The surface is generated from numerical integration of $\chi$. Note that $t$ is the dummy variable for integration in the $\zeta$-plane. \label{fig:surfstep}}
\end{figure}

Using the algorithms described in \S\ref{sec:riemann}, we generate the Riemann surface for $\chi$, and present the projection into $(\Re \zeta, \Im \zeta, \Re \chi)$-space in Fig.~\ref{fig:surfstep}. Recall the index notation of \eqref{chiexample}. In this case, the Riemann sheets are denoted by $\chi_{(k_1, k_2, k_3)}$ for the three branch points $\zeta = Z_i = \{ -b, -a, 0\}$. The main sheet, $\chi_{(0,0,0)}$, is shown at the very top of the surface in Fig.~\ref{fig:surfstep}, and also shown are two adjacent sheets, $\chi_{(0,0,1)}$ and $\chi_{(0,0,2)}$, achieved by passing through the logarithmic branch cut once or twice, respectively. The reader may compare the three-dimensional surface representation with the planar projection in Fig.~\ref{fig:stepstokes} to see how the branch cuts and individual Riemann sheets align. 

Examine now the lowest plane of Fig.~\ref{fig:surfstep}, where several representative steepest ascent/descent (or constant-phase) curves have been projected down from the surface. Locally setting $\zeta + a = r\e^{\i\vartheta}$, we find 
\begin{equation}
  \Arg \chi \sim \frac{5}{2}\vartheta - \frac{\pi}{2} \Rightarrow
  \begin{cases}
  \Im \chi = 0 & \text{when $\vartheta = \frac{2\pi}{5} \left(n + \frac{1}{2}\right)$,} \\
  \Re \chi = 0 & \text{when $\vartheta = \frac{2\pi}{5} n$,}
  \end{cases}
\end{equation}
for $n \in \mathbb{Z}$. Relating to Fig.~\ref{fig:surfstep}, we see dashed constant phase lines emerging from $\zeta = -a$ and forming two homoclinic orbits, separating the plane into three regions. Within the interior of the right-most homoclinic (formed by angles $\vartheta = \pm \pi/5$), the constant phase paths form closed orbits that become increasingly circular near $\zeta = 0$. The left-most homoclinic is formed by angles $\vartheta = \pm 3\pi/2$). Outside both homoclinic orbits, the constant phase paths also form close contours, approximately in the shape of a figure eight. Note that at the other critical point at $\zeta = -b$, which corresponds to the stagnation point of the step, $\chi' \neq 0$, and thus it is not a saddle point of the integrand.

\subsubsection{Only endpoint contributions} \label{sec:step_end}

\begin{figure}\centering
\includegraphics[scale=1]{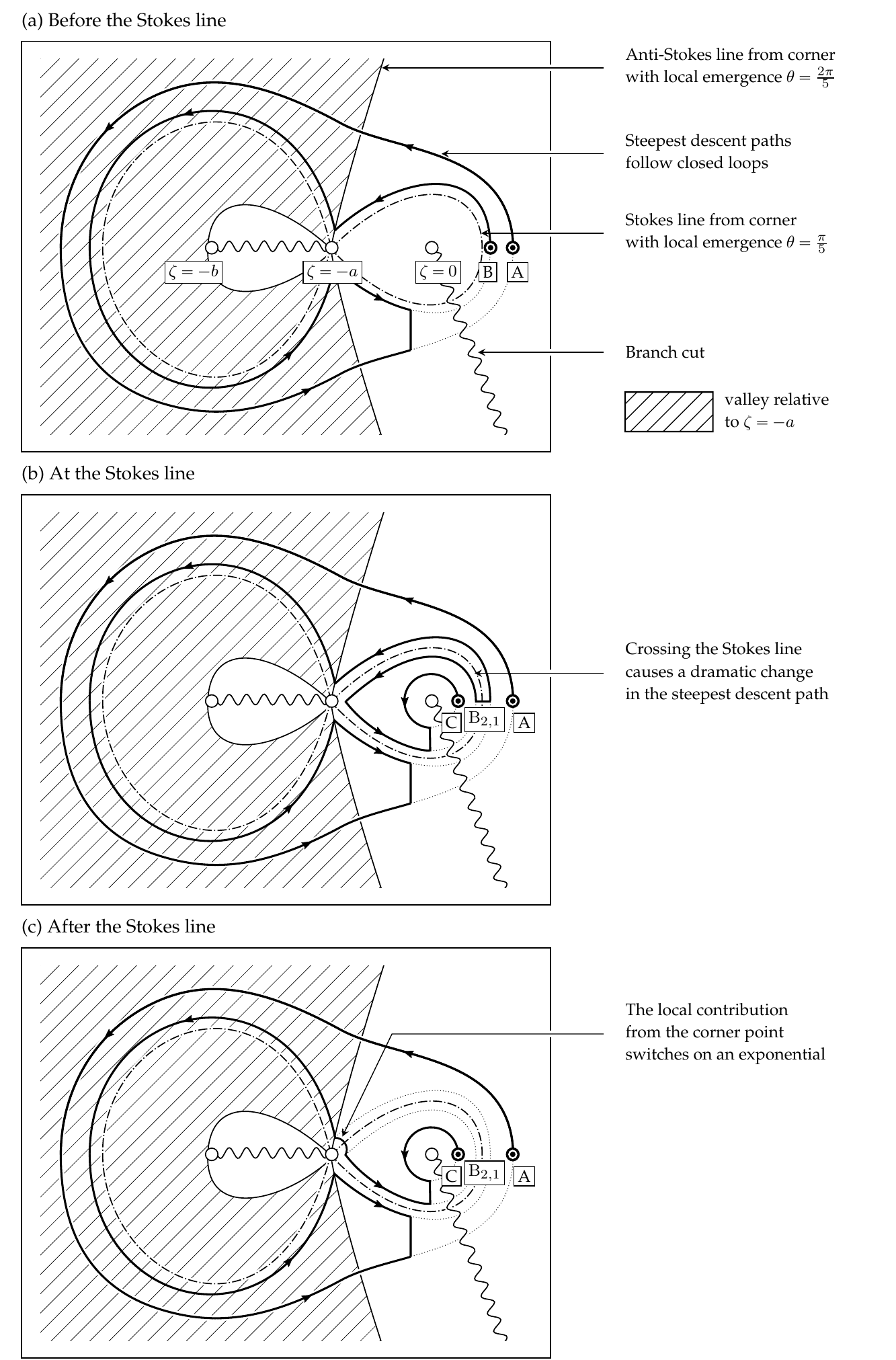}
\caption{Illustration of steepest descent deformations for the case of a rectangular step, shown in the complex $\zeta$-plane. Arrows indicate directions of descent. \label{fig:stepstokes}}
\end{figure}

We turn now to Fig.~\ref{fig:stepstokes}(a), and seek to deform the initial contour of integration in \eqref{eq:I_chi_step} when the two endpoints $t = \Ate$ and $t = \xi = \Bte$ are chosen sufficiently far upstream from the step so as to lie outside the right homoclinic. In the figure, we use $\zeta$ and $t$ interchangeably (the latter is a dummy variable of integration). 

Although the steepest descent path from each individual point forms a closed orbit in the $t$-plane, each full counterclockwise rotation about the figure eight trajectory is accompanied by a smaller value of $\Re(\chi)$. This is seen in the three-dimensional representation of Fig.~\ref{fig:surfstep}. Thus, the final deformed contour begins from A, and follows the trajectory counterclockwise, achieving smaller and smaller values of $\Re(\chi)$. In order to connect to the figure-eight contour joined to B, the connection is made within the valley where $\Re \chi \to -\infty$. Once done, the contour unwinds in a clockwise direction, and finally joins with B. Predictably, the result of this process is \eqref{Iend}, and thus $I$ is approximated by the two endpoint contributions. This yields the regular asymptotic expansion of the solution. 

Once we have understood this deformation process, it is simple to take the initial point $\Ate \to \infty$, that is, to be upstream infinity. We require the physical wave problem to impose a radiation condition such that the flow is uniform upstream and consequently, the only contribution to the integral at leading order is from an expansion about the point $\Bte$. 

\emph{We conclude that when sufficiently far upstream from the inclined-face of the step, there are no free-surface waves in the $\ep \to 0$ limit.}

\subsubsection{Crossing the Stokes line}

Now consider the deformation process when the initial contour of integration connects the upstream point, $t = \Ate$, with a point, $t = \Cte$, chosen sufficiently far downstream so as to lie within the right-most homoclinic orbit. This is shown in Fig.~\ref{fig:stepstokes}(b). We split the integration range in an analogous manner to \eqref{eq:I_steep_note}, with points $t = \Bte_1$ and $\Bte_2$, chosen to lie just on the exterior and interior, respectively, of the homoclinic orbit. The critical dividing line (the Stokes line), shown dash-dotted in the figure is the constant phase line from the step corner, $t = -a$. 

The steepest descent process then proceeds similarly as for the case of the surface-piercing ship: endpoint contributions are obtained from $\Ate$ and $\Cte$, and a saddle-point contribution is obtained from the corner of the step. Unlike the steepest descent paths for the ship, which tended towards either $t = 0$ or $t = \infty$, the steepest descent contours for the rectangular step form closed orbits, with each rotation diminishing the value of the integrand. This is shown most clearly in the three-dimensional surface of Fig.~\ref{fig:surfstep}. 

\emph{We conclude that when sufficiently far downstream from the rectangular step, the free-surface waves arise from switching-on the saddle-point contribution due to the corner of the step (the Stokes Phenomenon)}.

\subsection{The angled step in a channel} \label{sec:stepangle}

In both the rectangular ship and rectangular step, we have shown how the intersection of a Stokes line from the corner of the body with the free-surface accounts for the production of water waves. In more exotic geometries, multiple points in the body may produce multiple relevant Stokes lines. Bodies with smooth geometries will also produce waves as well---this does not invalidate our methods, and we return to discuss this in \S\ref{sec:discuss}. For now, let us consider the accompanying deformation of the Riemann surface in light of a change to the angle, $\pi\sigma$, of the step in Fig.~\ref{formstep}. 

Observe that Stokes lines correspond to steepest descent curves emerging from one of the critical points of the integrand. With $w_0$ in \eqref{eq:steepcrit} taken to be the point in question, then necessarily $\chi \to 0$ as the point is approached. Consider the stagnation point in the angled step, $\zeta = -b$. From \eqref{eq:I_chi_step}, we would require $\sigma < 1/3$ in order for $\chi \to 0$. This criterion led Chapman \& Vanden-Broeck~\cite{chapman_2006_exponential_asymptotics} to conclude that a Stokes line could only be produced by a point with an in-fluid angle greater than $2\pi/3$. Even at the proper angle, however, a Stokes line may exist, but it may not intersect the free surface (and thus makes no difference to the subsequent steepest descent paths).

Let us take for instance a shallow angled step with $\sigma = 1/10$. For this example, both $\zeta = -a$ and $\zeta = -b$ produce a relevant Stokes line. The Riemann surface for the step geometry is shown in Fig.~\ref{fig:angledstep}, and the picture reveals the remarkable complexity for even this moderately simple geometry. We note that the shallow angle of the step, as compared to the rectangular case, forces certain steepest descent contours to now travel past the branch cut associated with $\zeta = -b$, and move onto the sheet $\chi_{(-1, 0, 0)}$. 

The steepest descent process will go through three possible configurations, depending on the extent of which $\xi$ in \eqref{eq:I_chi_step} lies downstream from the step. If $\xi$ has not passed the first Stokes line from the corner, $\zeta = -a$, then only the end-point contribution is derived from the steepest descent process and there are no waves. Once $\xi$ passes the first Stokes line intersection from the corner, a single free-surface wave is switched on---as in the rectangular case. However, once $\xi$ passes the second Stokes line, a second wave is switched on. Thus, for this most downstream point, we would modify \eqref{IplusIexp} to
\begin{equation} 
   I \sim \Iend + I_{\textrm{exp, stag}} + I_{\textrm{exp, corner}} = \Iend + \left(\cint_{t = -b} + \cint_{t = -a}\right) \tilde{f}(t) \e^{\chi(t)/\ep} \, \de{t}.
 \end{equation} 
From the Riemann surface, notice that saddle from $t = -b$ lies lower than the one from $t = -a$. Therefore, the second free-surface wave will be exponentially subdominant to the first.

\afterpage{
\clearpage
\begin{landscape}
\thispagestyle{lscape}
\pagestyle{lscape}
\begin{figure}[hp]
\includegraphics{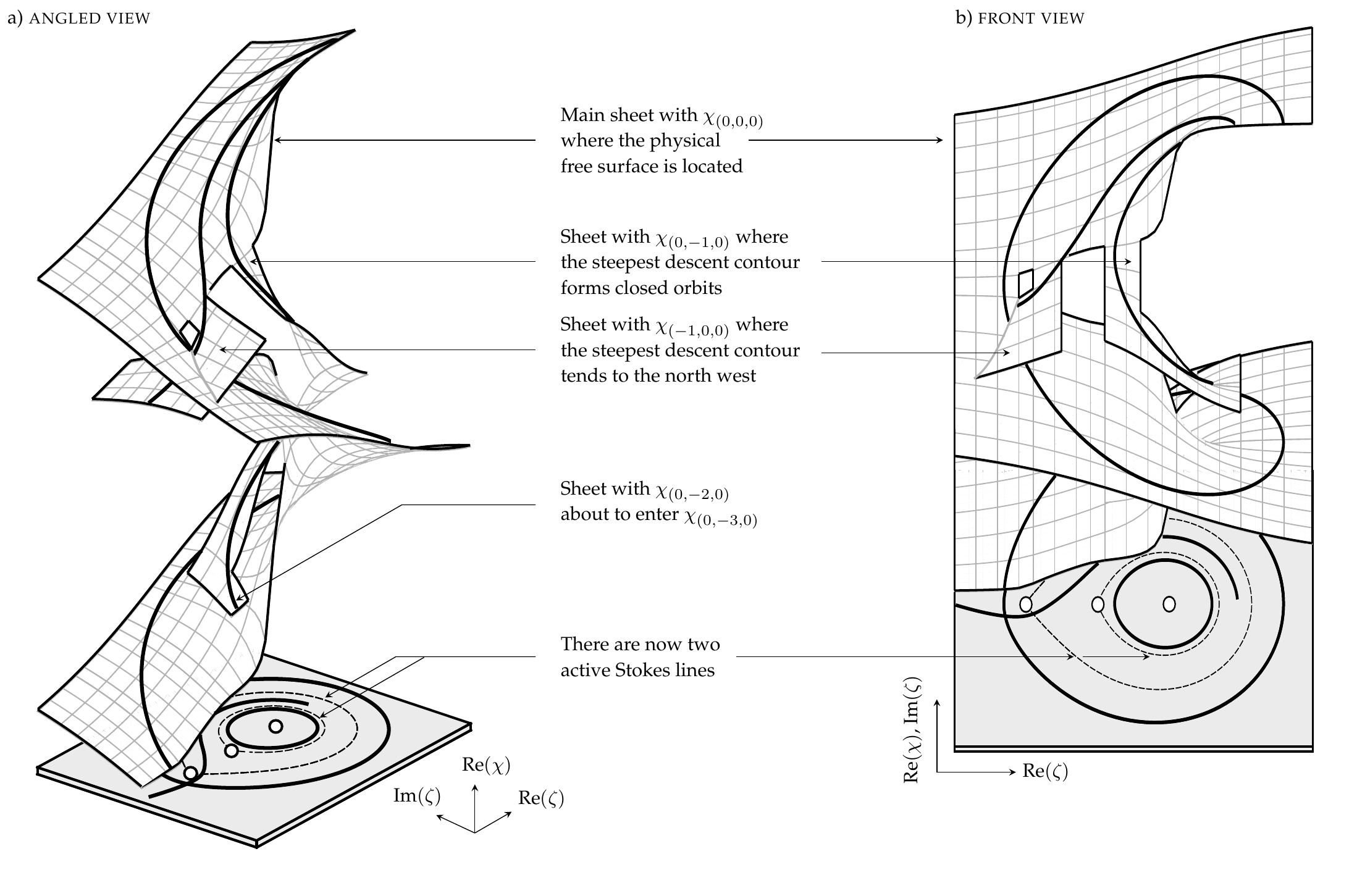}
\caption{The Riemann surface for the angled step with $\sigma = 1/10$, as projected into $(\Re \zeta, \Im \zeta, \Re \chi$)-space. The individual sheets are denoted by $\chi_{(k_1, k_2, k_3)}$ associated with the indicies, $k_i$ of $\zeta_i = \{-2, -1, 0\}$, for the step stagnation point, corner, and downstream infinity, respectively. Note that $t$ is the dummy variable for integration in the $\zeta$-plane. \label{fig:angledstep}}
\end{figure}
\end{landscape}
}

\section{Discussion} \label{sec:discuss}


The work in this paper was inspired through three principal motivations.

The first motivation was to return to the work of Tulin~\cite{tulin_1982_an_exact} and Tuck~\cite{tuck_1991_waveless_solutions,tuck_1991_ship-hydrodynamic_free-surface}, both of whom had proposed reduced models for free-surface flows with moving bodies. Tulin's model relied on a strategic trigonometric substitution inspired by Davies~\cite{davies_1951_the_theory}, while Tuck's model was less rigorous and more pedagogic in nature. Both models were united by their dependence on simplifying or outright removing the Hilbert transform, and also emphasized the importance of analytic continuation into the complex plane. Both models were also incomplete; thus the search for a complete model led to the companion work~\cite{trinh_tulinmodel_paper}, where a systematic reduction of the water-wave equations demonstrated why certain simplifications can be applied in the low-speed, $\ep \to 0$, limit. There, the analysis had depended on a general steepest descent argument, attributing surface waves to saddle-point contributions. Despite the relative simplicity of such arguments, the actual deformation process can be quite nontrivial, and in this paper, our objective was to demonstrate how this steepest descent analysis was performed for particular flow geometries. 

Our second motivation was to provide a powerful and visual methodology for the study of wave-structure interactions. The method establishes a correspondence between the physical flow with the topology of the Riemann surface that underlies paths of steepest descent. For instance, through the visualizations of Figs.~\ref{fig:surfship} and \ref{fig:surfstep}, we showed how to understand the effects of changing from flow past a surface-piercing ship, to flow over a step, simply by observing the associated changes in the  Riemann surfaces. Subtle geometrical changes can also be noted: for example in Fig.~\ref{fig:angledstep}, we changed the angle of the step, showing how the modification shifts paths of steepest descent onto previously untraveled areas of the Riemann surface.

Finally, our last motivation was to explore the various connections between the work here, with other research where exponential asymptotics is used to study free-surface flows. The successful application of exponential asymptotics to such problems has led to an exciting revival of classical problems in fluid mechanics~\cite{chapman_2006_exponential_asymptotics, trinh_2011_do_waveless}, but the methodology has always been limited by its relative complexity. There is a key simplification afforded by the study of \emph{gravity}-driven flow, and that is that Bernoulli's equation \eqref{govship} is only first-order in its derivative. Our integral methodology exploits this fact, and allows the solution to be written in the quasi-closed form \eqref{myint}, and then studied using the method of steepest descents. 

\emph{Is it then possible to consider other problems where a similar visual methodology is developed?} We are equally motivated by a need to better understand free-surface waves with the combined effects of gravity and surface tension. It is well known that in the case of gravity-capillary flows, the equivalent exponential asymptotics analysis is a great deal more complicated \cite{trinh_2013_new_gravity-capillary,trinh_2013a_new_gravity-capillary}. The difficulty with the inclusion of surface tension is that Bernoulli's equation in \eqref{bern} becomes of second order (due to the surface curvature terms), so the associated equations are not so easily studied using the method of steepest descents. Similarly, the study of time-dependent or three-dimensional problems using exponential asymptotics is still very much in its infancy \cite{chapman_2005,lustri_2014a,howls_1997_hyperasymptotics_for}.

\emph{What other geometries can be studied using this approach?} The reader may remark that, in reality, ships or channel topographies may not be piecewise-linear in shape as they are considered here. In our work, the generation of waves is attributed to critical points in the geometry (corners or stagnation points), but surely a smooth object will still create a wave pattern! In fact, our analysis does not stipulate that the body must contain singularities \emph{par se}, but rather the analytic continuation of $q$ and $\theta$.

Consider, for example, the replacement of the bottom topography in Fig.~\ref{formstep} by any streamline of the flow above the step. The free-surface waves must necessarily remain the same, and in fact, the steepest decents analysis previously applied will \emph{still} apply, as the (now-hidden) complex singularities of the step remain in place. There is no contradiction. Our work here also opens the possibility of studying flows past bodies with singularities weaker than the branch points found in the ship and step flows. In a forthcoming paper, we will study flows past bodies with singularities in the \emph{curvature} (rather than the angle), such as for the bulbous sterns considered in \cite{farrow_1995_further_studies}.





\begin{appendix}

\section{Governing equations} \label{sec:gov}

The general framework for the study of steady irrotational two-dimensional free-surface waves past a moving body is presented in the companion paper \cite{trinh_tulinmodel_paper}, and we also refer readers to the textbook by Vanden-Broeck~\cite{vb_book} for a more extensive review. Here, we summarize the basic equations.

We begin with a set-up of the three relevant coordinate systems. First, the physical plane with $z = x + \im y$ is chosen so that the body, which is moving at dimensional speed $U$, is fixed in a traveling frame of reference.  Second, in the complex potential plane with $w = \phi + \im \psi$, the free surface is assumed to lie at $\psi = 0$ and the fluid region in $\psi < 0$. In the case of the semi-infinite surface-piercing flow, the body lies on $\phi < 0$ and the free-surface on $\phi > 0$, and a stagnation point attachment is assumed at $\phi =0$. For channel flow, the fluid is confined within a finite strip, say $-\pi \leq \psi \leq 0$, with $-\infty < \phi < \infty$. The non-dimensional depth of the strip is chosen to be $\pi$ for convenience. 

The third coordinate system is required only for the case of channel flow. Here, an additional mapping
\begin{equation} \label{wtozeta}
  \zeta = \xi + \im \eta = \e^{-w},
\end{equation}
brings the fluid region to the upper half-$\zeta$-plane. The channel bottom now lies on $\xi < 0$, and the free surface on $\xi > 0$. For consistency of notation for the case of surface-piercing flows, we also consider the lower half-$\zeta$-plane using the identity map $\zeta = w$. The relevant coordinate systems are shown in Fig.~\ref{fig:outline} (for the ship) and \ref{formstep} (for the step).

A boundary integral formulation of the potential flow equations yields the following system of equations for $q$ and $\theta$ along the free surface: 
\begin{subequations} \label{govsys}
\begin{gather}
\ep q^2 \dd{q}{\phi} + \sin \theta = 0, \label{bern} \\
\log q = \frac{\j}{\pi} \dashint_{0}^\infty \frac{\theta(\xi')}{\xi' - \xi}  = \log q_s + \j \H[\theta](\xi) \label{bdint}.
\end{gather}
\end{subequations}

The first, \eqref{bern}, corresponds to Bernoulli's equation, and imposes a constant pressure condition on the surface. The second, \eqref{bdint}, involves a Cauchy principal value integral, and is equivalent to the imposition of Laplace's equation in the fluid region; it thus provides a second relationship between $q$ and $\theta$ that closes the system. In this latter equation, we have split the integral between the negative real axis (the body) and the positive real axis (the free surface), and defined the \emph{body function} $q_s$ and the Hilbert transform operator $\H$ according to
\refstepcounter{equation}\label{qsH}
\begin{equation} 
\log q_s = \frac{\j}{\pi} \int_{-\infty}^0 \frac{\theta(\xi')}{\xi' - \xi} \ \de{\xi'} \quad \text{and} \quad 
  \H[\theta](\xi) = \frac{1}{\pi} \dashint_{0}^{\infty}
  \frac{\theta(\xi')}{\xi' - \xi} \ \de{\xi'}. \tag{\theequation a,\,b}
\end{equation}
Above, we furthermore introduce the sign function
\begin{equation} \label{jsign}
    \j = \begin{cases}
    1 & \text{for surface-piercing flow,} \\
    -1 & \text{for channel flow,}
    \end{cases}
\end{equation}
which accounts for the fact that, in the surface-piercing geometry, the in-fluid-region lies to the right of the free-surface for increasing $\xi$, but to the left for channel flows. 

A given body geometry can be specified by providing the values of $\theta$ over $\xi < 0$ and then, by evaluating the integral (\ref{qsH}a), a body function $q_s$ is obtained, that encodes the appropriate geometry. Notice that when $\ep = 0$, $\theta = 0$ from \eqref{bern}, and thus $q = q_s$. This is the leading-order solution of a perturbative expansion $q = q_0 + \ep q_1 + \ldots$ and thus, $q_0 = q_s$. 

\section{Reduction of water wave equations to integral form} \label{sec:reduce}

For completeness, we summarize the main reduction of the water wave equations \eqref{govsys} to the integral form. The full details of this reduction can be found in \cite{trinh_tulinmodel_paper}. 

The first step is to analytically continue the free surface quantities $q(\phi, \psi = 0)$ and $\theta(\phi, \psi = 0)$ into the complex $\phi$ (or $\xi$ plane). We thus set $\phi \mapsto \phi_r + \i \phi_c$ and $\xi \mapsto \xi_r + \i \xi_c$. However, due to the correspondence between the (physical) complex velocity $w = \phi + \i \psi$ (with $\phi, \psi \in \mathbb{R}$) and the new \emph{complexified} $\phi$, we can relabel the analytic continuations to use $w$, with a similar relabeling for $\zeta$. Thus we write $q(\phi, \psi = 0) \mapsto q(w)$ and $\theta(\phi, \psi = 0) \mapsto \theta(w)$. This step of analytic continuation is explained in more detail in \cite{trinh_2014_the_wake,trinh_2015_exponential_asymptotics}, and the reader is also encouraged to see the visualization in Fig.~2 of \cite{trinh_2013a_new_gravity-capillary}. Under this analytic continuation, the boundary integral equation \eqref{bdint} becomes
\begin{equation} \label{bdintcomplex}
  \log q + \i \j \theta = \log q_s + \j \bH[\theta] \quad \text{where} \quad
  \bH[\theta](\zeta) = \frac{1}{\pi} \int_{0}^{\infty} 
\frac{\theta(\xi')}{\xi' - \zeta} \ \de{\xi'},
\end{equation}
for analytic continuation into the upper half-$w$ or $\zeta$-plane. 

Now the solutions, $q$ and $\theta$ are expressed in terms of a regular series expansion with an error term, \emph{i.e.} $q = q_r + \bq$, $\theta = \theta_r + \bt$ as in \eqref{qtseries}. Note that the leading-order $q_0$ is given by $q_s$ in (\ref{qsH}a) and $\theta_0 = 0$, known as the rigid-body flow. The combination of Bernoulli's equation \eqref{bern} and the boundary integral equation \eqref{bdintcomplex} gives the result of \cite{trinh_tulinmodel_paper}: 

\begin{result}[Reduced integro-differential model] \label{result:integro}
Linearizing the water wave equations about a regular series expansion truncated at $N$ terms gives the following integro-differential equation for the perturbation, 
\begin{subequations} \label{simpsys}
\begin{equation} \label{simpsyseq}
\ep \bq' + \biggl[ \chi'(w) + \ep P_1'(w) + \Oh(\ep^2) \biggr]\bq 
= R(w; \bH[\bt]) + \Oh(\bt^2, \bq^2).
\end{equation}
where
\begin{gather}
\chi(w) = \int_{w_0}^w \frac{\im \j}{q_0^3(\varphi)} \, \de{\varphi}, \label{chi} \\
\e^{-P_1(w)} =  q_0^2(w^*) Q(w) =  \left[ \Lambda \frac{q_0^2(w^*)}{q_0^2(w)}\right]  
\exp\left( 3\i \j \int_{w^*}^w \frac{q_1(\varphi)}{q_0^4(\varphi)} \, \de{\varphi}\right)
\label{P1} \\
R(w; \bH[\bt]) = -\Eb + \im \bH[\bt]\frac{\cos\theta_r}{q_r^2}, \label{Rfunc}
\end{gather}
and the error term, $\Eb$, represents the error in Bernoulli's equation, and is given by
\begin{equation} \label{Ebern}
  \Eb = \ep q_r' + \frac{\sin\theta_r}{q_r^2}.
\end{equation}
The initial point of integration, $w_0$, in \eqref{chi} is typically chosen at an (integrable) singularity of $q_0$, given by (\ref{qsH}a). The point of integration, $w^*$, can be chosen anywhere the integral in \eqref{P1} is defined, and only changes the constant of integration, $\Lambda$. 
\end{subequations}
\end{result}

The solution of the first-order differential equation \eqref{simpsyseq} can then be written as the integral in \eqref{myint}. The initial point of integration is taken so as to satisfy the boundary or radiation conditions. For the case of the ship \eqref{eq:q0ship}, we set $\ws = 0$ (for the stagnation point), while for the case of the step, we set $\ws = -\infty$ (for a wave-free upstream flow). 

As discussed in \cite{trinh_tulinmodel_paper}, different choices for $R$ can be used in order to obtain different approximations to the waves. A particularly convenient and often very accurate choice is the two-term approximation,
\begin{equation} \label{Rtest}
R(w; \bH[\bt]) \sim \ep^2 \left(-\frac{5\im\j q_1^2}{2 q_0^4} + q_1' + 2\im \frac{\bH[\theta_1]q_1}{q_0^3}\right),  
\end{equation}
which was used to illustrate the numerical example of Fig.~\ref{fig:outline}. A comparison of different choices for $R$ is presented in Table~2 of \cite{trinh_tulinmodel_paper}. 
\end{appendix}


\bibliographystyle{plainnat}

\providecommand{\noopsort}[1]{}

\end{document}